\documentclass{ws-p8-50x6-00}

\begin{document}

\def\p{\phi}\def\P{\Phi}\def\a{\alpha}\def\e{\epsilon}
\def\be{\begin{equation}}\def\ee{\end{equation}}\def\l{\label}
\def\0{\setcounter{equation}{0}}\def\b{\beta}\def\S{\Sigma}\def\C{\cite}
\def\r{\ref}\def\ba{\begin{eqnarray}}\def\ea{\end{eqnarray}}
\def\n{\nonumber}\def\R{\rho}\def\X{\Xi}\def\x{\xi}\def\la{\lambda}
\def\d{\delta}\def\s{\sigma}\def\f{\frac}\def\D{\Delta}\def\pa{\partial}
\def\Th{\Theta}\def\o{\omega}\def\O{\Omega}\def\th{\theta}\def\ga{\gamma}
\def\Ga{\Gamma}\def\t{\times}\def\h{\hat}\def\rar{\rightarrow}
\def\vp{\varphi}\def\inf{\infty}\def\le{\left}\def\ri{\right}
\def\foot{\footnote}\def\ve{\varepsilon}\def\N{\bar{n}(s)}\def\cS{{\cal S}}
\def\k{\kappa}\def\sq{\sqrt{s}}\def\bx{{\bf x}}\def\La{\Lambda}
\def\bb{{\bf b}}\def\bq{{\bf q}}\def\cp{{\cal P}}\def\tg{\tilde{g}}
\def\cf{{\cal F}}\def\bN{{\bf N}}\def\Re{{\rm Re}}\def\Im{{\rm Im}}
\def\K{{\bf K}}\def\F{{\bf Fig}}\def\S{{\bf S}}\def\e{\varepsilon}

\title{Why the Very High Multiplicity events are rare}

\author{J.Manjavidze}

\address{Joint Inst. for Nucl. Res., Dubna, Russia and Inst. Phys., Tbilisi, Georgia\\
E-mail: joseph@nusun.jinr.ru}

\author{A.Sissakian}

\address{Joint Inst. for Nucl. Res., Dubna, Russia\\
E-mail: sisakian@jinr.ru}

\maketitle

\abstracts{The possibility to suppress the nonperturbative effects
choosing the vary high multiplicity final state is discussed. The
theoretical uncertainties and the experimental observable
consequence of this choice are discussed.}

{\bf\large 1.} The topic of present report is to show the reason
{\it why the experiments with Very High Multiplicity (VHM) final
state of hadrons may be important}. Actually, we imply here that
the investigations in VHM region may give the information which is
not attainable in other experiments.

One of the reasons of necessity to examine the VHM processes is
connected with following fundamental question:

$\star$ {\it Why the VHM events are so rear}?\\ In other words, we
offer look into the question: why the mean multiplicity in the
hadron processes is so small ($\sim \ln^2 s$) in comparison with
the multiplicity threshold value ($\sim \sqrt s/m_\pi$).

Present paper is based on the described in \C{1} idea that the
very hot cup of tea cooling in the very cold room is the analogy
of VHM process. Intuitively we know that in this case the cooling
process should proceed quickly.

In field-theoretical terms the "fastness" of process means that
the decay of virtual coloured partons on secondaries should
prevail over its dispersion from interaction zone. This becomes
possible if and only if the parton virtuality is high enough: if
the "virtuality" of constituent is $|q|^2$ then the life time of
such object $\sim1/|q|$ \C{rep}. But this means that the process
should be hard.

In addition, the produced particle in VHM kinematics have the low
energies. Last one singles out VHM processes from other hard
processes (i.e., from hadron production in the
$e^+e^-$-annihilation process, or DIS processes, etc.).

\begin{figure}[t]
\epsfxsize=15pc \figurebox{20pc}{15pc}{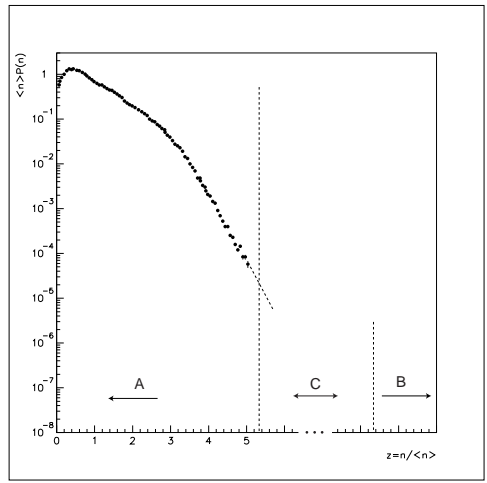}
\caption{Multiplicity distribution. Points are E735 (Tevatron)
data. {\bf C} is the VHM region. \label{fig:radish}} \vskip -1cm
\end{figure}

Fig.1 defines the range of VHM domain. But it should be stressed
that, in definite sense, this determination is conditional and
depends on the model.

Yet, we will assume that the multiperipheral kinematics governs
the dynamics in the range {\bf A}. The range {\bf B} assumes that
the particle momentum is less then its mass and for this reason
the dynamics did not play any role in it. We will assume that
$n<<n_{max}=\sqrt{s}/m_\pi .$ So, we would interested in the range
{\bf C}, where the "trivial" multiperipheral picture do not work,
but the dynamics is still important.

{\bf\large 2.} The standard point of view on hadron dynamics based
on the fact that it is "soft", the diffraction radii increase with
energy and the mean transverse momentum of secondaries is
approximately constant. This picture is described, actually
postulating, by, so called, {\it soft Pomeron} model \C{2}.

This soft processes may be approximated by the very spatial class
of ("ladder") Feynman diagrams of perturbative quantum
chromodynamics (pQCD). This is so called "{\it BFKL Pomeron}"
approach \C{3}. Both model absorb the condition that the dynamics
of main multiple production processes is restricted. In the {\it
soft Pomeron} case this constrains are the "hidden conservation
laws" as the consequence of underlying non-Abelian gauge symmetry.
The LLA ideology is used in the {\it BFKL Pomeron} case.

The constrains restrict production dynamics and the mean
multiplicity in that way constraint processes must be small, i.e.
$n<<n_{max}$ should be important.

So, this constrains should be suppress to have the VHM final
state. Notice also that the long-range constrains suppression
assume change of mechanism of particles production in the VHM
region {\bf C}. Discussed scenario would realized if (i) the hard
processes may prevail over the soft ones  and if (ii) the hard
processes may lead to the "fast" production.

Formal prove contains following steps \C{1,4}. So, it can be shown
that if the "final-state" interaction are excluded then the soft
processes may provide only following asymptotics: $\s^s_n<
O(e^{-n}).$ On other hand, the hard process gives following
asymptotic estimation: $\s^h_n=O(e^{-n}).$ Therefore, one always
may find such energy $\sqrt s$ and multiplicity $n$ that
$\s^h_n>>\s^s_n~{\rm for}~n<<\sqrt{s}/m_\pi.$ This proves (i).

The qualitative argument in favor of (ii) follows from the
Eurenfest-Kac model \C{5} of the irreversibility phenomenon. It
shows that if initial state is {\rm far} from final one then the
system goes to equilibrium {\it as fast is possible}, see absence
of fluctuations on the early stage of the process on Fig.2.

\begin{figure}[t]
\epsfxsize=15pc \figurebox{20pc}{15pc}{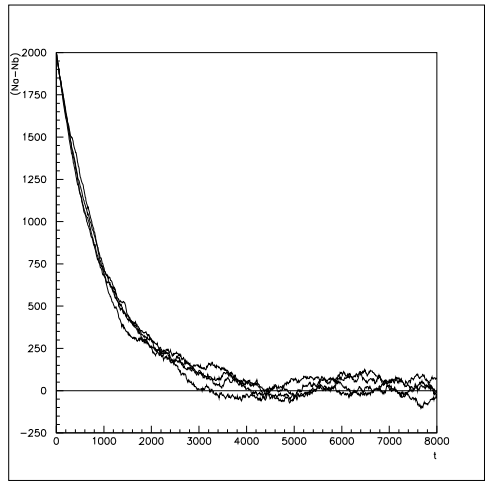}
\caption{Particles production in the Markovian process
\label{fig:radish}}\vskip -0.8cm
\end{figure}

The formal prove of supposition (ii) uses the KNO-scaling form of
the particles number distribution in the pQCD jets \C{6}.

{\bf\large 3.} We know well that it is necessary to have definite
proportional to the particle energy $\e$ "formation length" for
particle production in the hadron-ion collision. For this reason
the mean multiplicity in the hadron-ion collision is comparatively
small, it is proportional to the radii of ion.

To have the VHM final state one should have enormous reproduction.
So, the VHM final state may produced in hadron-ion collisions if
the "formation length" is always smaller than the ion radii. In
this case all nucleon work in the process of particles production.
So, the VHM states production is the "volume effect": the
measurable should dependent on the whole number of the ion
constituents. The detailed investigation of the VHM states in the
ion-ion collision is intensively performed.

It was shown that the VHM processes should be hard. Then, one can
say that such process is freely evolves. It is the important
observation since allows to conclude that the final state reach in
this case {\it the equilibrium condition} \C{1,7}.

Observed hadron system contains definite number of constrains, but
not enough to suppress the particle production completely. This
conclusion follows from existence of large multiplicity
fluctuations, see Fig.1. But in the domain {\bf A} the constrains
are important and mean multiplicity is small in comparison with
$n_{max}$. We investigate this conclusion theoretically and the
result is following: in the region {\bf B} the system without fail
is equilibrium.

We found the experimentally measurable characteristic, which
measures the "equilibrium phenomena". For instance, absence of
energy correlations mean the thermal equilibrium. We would like to
underline that the condition $R=|K_3|^{2/3}/|K_2|<1$ is the
necessary and sufficient condition of thermalization. The quantity
$K_l$ is the $l$-particle central energy correlator, $l=2,3,...$.

Fig.3 shows the Monte Carlo simulation of the ration $R(n,s)$ for
domain {\bf A} using PYTHIA generator of events. Absence of
thermalization in the domain {\bf A} is natural since PYTHIA
resembles the Regge model. At the same time, it is simple to show
that in the region {\bf B} the system without fail should reach
the thermal equilibrium: $R\sim 1/n$ in this domain.

\begin{figure}[t]
\epsfxsize=18pc \figurebox{20pc}{15pc}{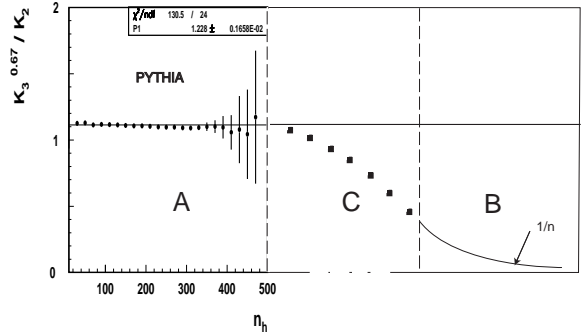}
\caption{Ratio $R(n,s)=\le|K_3\ri|^{2/3}/\le|K_2\ri|$ vs.
multiplicity. PYTHIA prediction in {\bf A}. \label{fig:radish}}
\vskip -1cm
\end{figure}

We can conclude that VHM processes

(i) allows to investigate the hadron dynamics beyond standard
(multiperipheral) kinematics,

(ii) are hard and the influence of the hidden constrains do not
play the important role,

(iii) can not be described using ordinary LLA ideology \C{9},

(iv) allows to reach the thermal equilibrium,

We would like to add at the very end that, as follows from above
conclusion, the VHM processes are the source of $dense$, $cold$
and $equilibrium$ locally coloured state (plasma).

\section*{Acknowledgments}
Authors would like to take the opportunity to thank E.Levin,
L.Lipatov, V.Matveev, V.Nikitin for important discussions. Special
thanks to ATLAS (LHC, CERN) physics community for stimulating
interest to the VHM problem.

\end{document}